\documentclass[useAMS,usenatbib]{mn2e}

 \usepackage{times}

\def\pcm3{{\rm\thinspace cm^{-3}}}

\def\contcaption{\@conttrue\SFB@caption\@captype}

\def\n_h{{\rm n_{H}}}

\def\NH1{{$N_{\rm HI}~$}}

\def\ga{{\rm\thinspace gauss}}

\def\approxlt{\mathrel{\hbox{\rlap{\lower .5ex \hbox {$\sim$}}
        \raise .15 ex \hbox{$<$}}}}
\def\approxgt{\mathrel{\hbox{\rlap{\lower .5ex \hbox {$\sim$}}
        \raise .15 ex \hbox{$>$}}}}

\def\la{\mathrel{\hbox{\rlap{\hbox{\lower4pt\hbox{$\sim$}}}\hbox{$<$}}}}
\def\ga{\mathrel{\hbox{\rlap{\hbox{\lower4pt\hbox{$\sim$}}}\hbox{$>$}}}}
\newbox\grsign \setbox\grsign=\hbox{$>$} \newdimen\grdimen
\grdimen=\ht\grsign
\newbox\simlessbox \newbox\simgreatbox \newbox\simpropbox
\setbox\simgreatbox=\hbox{\raise.5ex\hbox{$>$}\llap
     {\lower.5ex\hbox{$\sim$}}}\ht1=\grdimen\dp1=0pt
\setbox\simlessbox=\hbox{\raise.5ex\hbox{$<$}\llap
     {\lower.5ex\hbox{$\sim$}}}\ht2=\grdimen\dp2=0pt
\setbox\simpropbox=\hbox{\raise.5ex\hbox{$\propto$}\llap
     {\lower.5ex\hbox{$\sim$}}}\ht2=\grdimen\dp2=0pt
\def\simgreat{\mathrel{\copy\simgreatbox}}
\def\simless{\mathrel{\copy\simlessbox}}

\title[]{Photospheric phosphorus in the FUSE spectra of GD71 and two similar DA white dwarfs}
\author[P. D. Dobbie et al.]{P. D. Dobbie$^{1}$\thanks{E-mail:
pdd@star.le.ac.uk} M.A. Barstow$^{1}$ I. Hubeny$^{2}$ J.B. Holberg$^{3}$ M.R. Burleigh$^{1}$ A.E. Forbes$^{1}$  \\
$^{1}$Department of Physics and Astronomy, University of Leicester, University Road, Leicester LE1 7RH, UK\\
$^{2}$Dept of Astronomy and Steward Observatory, University of Arizona, Tucson, AZ 85721, USA\\
$^{3}$Lunar and Planetary Laboratory, Gould-Simpson Building, University of Arizona, Tucson, AZ 85721, USA \\
}
\begin{document}

\date{Accepted 1988 December 15. Received 1988 December 14; in original form 1988 October 11}

\pagerange{\pageref{firstpage}--\pageref{lastpage}} \pubyear{2002}

\maketitle

\label{firstpage}

\begin{abstract}
We report the detection, from FUSE data, of phosphorus in the atmospheres of GD71 and two 
similar DA white dwarfs. This is the first detection of a trace metal in the photosphere of 
the spectrophotometric standard star GD71. Collectively, these objects represent the coolest 
DA white dwarfs in which photospheric phosphorus has been observed. We use a grid of homogeneous 
non-LTE synthetic spectra to measure abundances of [P/H]=$-8.57^{+0.09}_{-0.13}$, 
-8.70$^{+0.23}_{-0.37}$ and -8.36$^{+0.14}_{-0.19}$ in GD71, RE J1918+595 and RE J0605-482
respectively. At the observed level we find phosphorus has no significant impact on the overall 
energy distribution of GD71. We explore possible mechanisms responsible for the presence of 
this element in these stars, concluding that the most likely is an interplay between 
radiative levitation and gravitational settling but possibly modified by weak mass loss.

\end{abstract}


\section{Introduction}

The compositions of the atmospheres of white dwarfs are observed to be dominated by the lightest elements,
H (DAs) or He (DO/DBs), as predicted several decades ago by theory (Schatzman 1958). Unimpeded,
the high surface gravities of these objects cause heavier elements to settle out on timescales of mere days
(e.g. Dupuis et al. 1993). Nevertheless, ultraviolet observations have revealed trace quantities of metals in the 
photospheres of a large number of hot white dwarfs (T$_{\rm eff}\simgreat40000$K), e.g. C, Fe and
Ni in G191-B2B (Bruhweiler \& Kondo 1983, Vennes et al. 1992, Holberg et al. 1994) and
a handful of cooler degenerates e.g. Si in Wolf 1346 (T$_{\rm eff}\approx20000$K; Holberg et al. 1996). 
Furthermore, high resolution optical spectroscopic observations have shown that $\sim$20\% of DAs cooler
than 10000K are DAZ stars, that is they contain observable quantities of Ca, Mg, and Fe (Zuckerman et al.
2003). Clearly, there are processes operating in the photospheres of these objects which can compete against 
gravitational settling. 

Several theoretical studies have shown that the transfer of net upward momentum from the intense radiation 
field to heavy elements via their bound-bound transitions can prevent small but detectable quantities 
of metals from sinking out of the photospheres of white dwarfs with T$_{\rm eff}$$\simgreat$20000K(e.g.
Vauclair, Vauclair \& Greenstein 1979). A number of these investigations have made relatively precise
predictions regarding the abundances of heavy elements as a function of surface gravity and effective temperature 
(e.g Chayer et al. 1995a, Schuh 2005). Indeed, these calculations are able to qualitatively reproduce some of 
the more general trends, affirming radiative levitation as an influential mechanism in preventing gravitational 
settling in the atmospheres of these hotter stars. For example, they successfully replicate the observed sharp 
decline in the abundance of Fe at T$_{\rm eff}$$\approx$50000K and the observed decrease with T$_{\rm eff}$ 
in the overall metal content of white dwarf photospheres. Further, in accord with these calculations the abundances
of elements such as C and N appear to be largely independent of effective temperature at T$_{\rm eff}$$\simgreat
$50000K (see Barstow et al. 2003).

However, at a more quantitative level, there are significant discrepancies between observed heavy element 
abundances and the predictions of equilibrium radiative levitation theory. On theoretical grounds, stars with 
comparable 
effective temperatures and surface gravities are expected to have similar photospheric abundances, but in 
reality can show radically different compositions. For example, while the observed energy distribution of 
GD153 is consistent with that of a pure-H atmosphere, GD394 and RE1614-085 contain significant quantities of 
silicon ([Si/H]=-5.1$\pm$0.1) and nitrogen ([N/H]=-3.6$\pm$0.1) respectively, despite all three objects having
T$_{\rm eff}$$\approx$38000K and log g$\approx$7.8 (Holberg et al. 1997). It has often been argued that these 
anomalies indicate other processes (e.g. accretion or massloss), act in conjunction with radiative 
levitation, at least in some stars, to dictate photospheric composition.

Our survey of the abundances of C,N,O,Si,Fe and Ni in the photospheres of 25 hot DA white dwarfs is 
the most comprehensive undertaken to date and unearthed a number of interesting results (see Barstow 
et al. 2003). However, it was based largely on HST STIS data, a sample that cannot now be enlarged
due to instrument failure. To gain a more complete understanding of the relative influence of each of these 
physical processes it is important to continue enlarging the sample of stars for which robust measurements
and limits on the abundances of photospheric metals are available. Further, it is particularly important 
to increase the number of such stars in effective temperature ranges where few objects have previously been 
studied (e.g. 20000K$\simless$T$_{\rm eff}$$\simless$40000K). Over the last few years the Far Ultraviolet 
Spectroscopic Explorer (FUSE) has observed a large number of white dwarfs. This has led to the provision of 
a collection of homogeneous spectroscopic data covering $\lambda$$\approx$900-1200\AA\, with generally moderate 
to good signal-to-noise, for $\simgreat$100 DAs. Much of this data is now in the public domain and available
for download from the Multimission Archive at Space Telescope (MAST). These FUSE observations offer the 
benefit over IUE and HST data of access to resonance lines of P and S. Thus photospheric
abundance surveys can now be expanded to include these two additional elements. 

GD71 (WD0459+158), first catalogued by Giclas et al. (1965), is a bright nearby hot DA white dwarf lying 
virtually in the middle of the aforementioned temperature range (T$_{\rm eff}$$\approx$32000K).
It has been studied extensively at optical, UV, EUV and near-IR wavelengths and is widely used as a 
photometric and spectrophotometric calibration star (e.g. Landolt 1992, Bohlin et al. 2001). Within their respective statistical uncertainties, the International 
Ultraviolet Explorer (IUE) and Extreme Ultraviolet Explorer (EUVE) spectra of GD71 are entirely consistent 
with a pure-H atmosphere and to date no heavy elements have been detected in its photosphere (Barstow 
et al. 1997, Holberg et al. 1998). Indeed, for spectrophotometric calibration purposes it has been assumed to have
a pure-H composition (e.g. Bohlin 1996). However, we have chosen to examine this assumption more closely, since 1) 
radiative levitation theory predicts the presence of small but nevertheless detectable quantities of photospheric
metals down to T$_{\rm eff}$$\approx$20000K (e.g. Si, Al  and P),  2) to date only a handful of objects have been 
studied in detail in the range 20000K$\simless$T$_{\rm eff}$$\simless$40000K and 3) there exists a FUSE spectrum with 
good S/N ($\sim$30) for this white dwarf.

\begin{table}
 \centering
 \begin{minipage}{125mm}
\caption{The log of the FUSE observation of GD71.}
\begin{tabular}{@{}cccccc@{}}
\hline
Observation ID   &   Aperture/mode    &   Start date   &  Exp. time (secs.) \\
 \hline
P2041701000 & LWRS/TTAG & 00-Nov-04 & 13928 \\ 
\hline
\end{tabular}
\end{minipage}
\end{table} 

In the current work we present an analysis of the FUSE spectrum of GD71.
We identify a number of spectral features attributable to the interstellar medium. In addition, 
we detect high ionization P features which we argue arise in the stellar photosphere. We 
use a grid of non-LTE model atmospheres to determine [P/H]. Finally, we discuss our findings 
in the context of radiative levitation theory and the processes of accretion and mass-loss.

\section[]{Observations}
\subsection{A FUSE observation of GD71}

An observation of GD71 was obtained with FUSE operating 
in TTAG mode and in the LWRS configuration on 2000/11/04. We have acquired the relevant raw data 
products from the Multi-mission Archive at the Space Telescope Science Institute (MAST). A summary of 
this observation is presented in Table 1.

Several detailed descriptions of the FUSE instrumentation already exist in the literature (e.g. Green, 
Wilkinson \& Friedman 1994), so we only give a brief description of the 
features most relevant to the current analysis. The spectrometer consists of four separate optical 
channels. To maximize the throughput, it is important that all four are properly 
illuminated by the target. However, it has proved difficult to maintain optimal alignment for the 
duration of an observation due to in-orbit movement in the mirrors and gratings. Consequently, most 
observations have been obtained through
the large square aperture (30''x30'') of the LWRS configuration which limits the spectral resolution to
between R=10000-20000. 

We have processed the raw data with a recent version of CALFUSE (v3.0). The pipeline procedure
nominally flux and wavelength calibrates the data, correcting for geometric distortions and 
flagging as low quality data obtained on deadspot areas of the detectors. However, this version of the 
software does not correct for the impact on the count rates of the ``worm'', a strip of flux attenuated 
by up to 50\%, running in the dispersion direction of the spectra. An inspection of the count rate
plots for each channel produced by the pipeline suggests that GD71 remained comfortably within the 
LWRS apertures throughout the duration of this observation. Furthermore, an examination of the individual 
extracted 1D spectra indicates that the ``worm'' only affected LiF 1b data significantly.
 
Prior to coadding the individual datasets to obtain a single FUV spectrum of GD71 with optimal S/N, 
it was necessary to account for drifts in the wavelength scale between the subintegrations of the 
observation. The exposures for each segment were cross correlated by running the
the IDL routine CROSS-CORRELATE on the regions which included the most distinct absorption features 
(e.g. the NI triplet at 1134\AA\ for LiF 1b and Lif 2a). Small wavelength corrections were calculated and applied 
to the data. The alignments of the exposures for each segment were examined by eye, and, where improvement 
was necessary or possible (e.g. where geocoronal emission had impacted on the results of CROSS-CORRELATE 
algorithm), further small shifts applied manually to their wavelength scales. A co-added dataset for each 
segment was then produced where each subintegration was weighted according to exposure time. Any data found 
to be significantly affected by the worm were discarded. As they generally have the lowest S/N and the 
poorest wavelength solution, data from the edges of each spectrum were also rejected. Sections of these 
datasets which include strong absorption lines were cross correlated and any linear wavelength shifts 
with respect to the LiF 1a segment calculated and
applied to each. However, the FUSE spectrum of GD71 contains no strong absorption features in the ranges 
995-1010\AA\ and 1090-1110\AA. Therefore the measured wavelengths 
of sharp line features arising from species HI, CII, OI, and NI, which were assumed to have a common 
interstellar origin, were used to apply a fine adjustment to the wavelength scales of the SiC 1b, SiC 2a, LiF 1b 
and LiF 2a segments. 
Finally, the spectra were resampled onto a single wavelength scale with 0.04\AA\ binning and coadded,
weighting each by the inverse of the statistical variance of the data determined over intervals of
 20\AA. 

\subsection{Absorption features in the co-added FUV spectrum of GD71}

\begin{table}
\centering
\begin{minipage}{78mm}
\caption{Summary details of the main absorption features identified in the coadded FUSE spectrum of 
the white dwarf GD71}
\begin{tabular}{@{}lrrrrrr@{}}
\hline
Ion    & \multicolumn{1}{|c|}{Lab}    &   \multicolumn{1}{|c|}{Obs}     &  \multicolumn{1}{|c|}{v}        &   \multicolumn{1}{|c|}{$\Delta$v}   &   \multicolumn{1}{|c|}{E.W.}   &   \multicolumn{1}{|c|}{$\Delta$E.W.} \\
       & \multicolumn{2}{|c|}{\AA}   &   \multicolumn{2}{|c|}{kms$^{-1}$} & \multicolumn{2}{|c|}{m\AA} \\
\hline
\multicolumn{7}{|c|}{Interstellar} \\
HI     & 917.181  & 917.196    &   4.9 &  4.9 & 168 & 19\\
HI     & 918.129  & 918.130    &   0.3 &  4.1 & 135 & 16\\
HI     & 919.351  & 919.356    &   1.6 &  4.9 & 156 & 11\\
HI     & 920.963  & 920.959    &  -1.3 &  4.6 & 180 & 11\\
HI     & 923.150  & 923.152    &   0.6 &  8.3 & 194 & 12\\
HI     & 926.226  & 926.234    &   2.6 &  2.3 & 175 & 14\\
HI     & 930.748  & 930.728    &  -6.4 &  1.3 & 161 & 14\\
HI     & 937.803  & 937.780    &  -7.4 &  1.0 & 151 & 13\\
HI     & 949.743  & 949.747    &   1.3 &  1.3 & 175 & 13\\
OI     & 976.448  & 976.421    &  -8.3 &  1.5 &  10 & 4 \\ 
CIII   & 977.020  & 977.013    &  -2.1 &  1.5 &  41 & 6\\
NIII$^{\dag}$   & 989.799  & 989.838    &  11.8 &  3.9 & 31 & 4 \\
SiII$^{\dag}$   & 989.873  & 989.838    & -10.6 &  3.9 & 31 & 4\\
CII    & 1036.337 & 1036.334   &  -0.9 &  0.6 & 56 & 2\\
OI     & 1039.230 & 1039.242   &   3.5 &  1.5 & 11 & 2\\
NII    & 1083.990 & 1084.016   &   7.2 &  1.0 & 28 & 4\\
NI$^{\dag\dag}$     & 1134.165 & 1134.155  &  -2.6 &  5.7 &  4 & 2\\
NI$^{\dag\dag}$     & 1134.415 & 1134.426  &   2.9 &  1.8 & 17 & 3\\
NI$^{\dag\dag}$     & 1134.980 & 1134.997  &   4.5 &  1.2 & 15 & 2\\
FeII                & 1144.938 & 1144.948  &   2.6 &  3.7 &  4 & 2 \\  
\multicolumn{7}{|c|}{Photospheric} \\
PIII   & 1003.600 & 1003.615   &   4.5 & 5.1  &  5 & 2\\
PV     & 1117.977 & 1117.985   &   2.1 & 2.8  & 13 & 3\\
PV     & 1128.008 & 1128.020   &   3.2 & 1.7  &  4 & 2\\
\hline
\end{tabular}

{\small $^{\dag}$ Blended $?$} \\
{\small $^{\dag\dag}$ Some contamination from NI geocoronal emission} \\

\end{minipage}
\end{table}

The co-added FUSE spectrum has been run through a series of custom 
written scripts to (1) normalize the data, (2) flag probable absorption features, (3) fit Gaussians 
or where appropriate multiple Gaussians to determine central 
wavelengths and equivalent widths of features and (4)  assign each feature a likely identification
based on the measured radial velocity and the ionization state of the proposed progenitor species. 
The results of this process are summarised in Table 2. We note that the 1-$\sigma$ errors in 
central wavelengths, derived from a $\chi^{2}$ fitting process, underestimate the true 
uncertainties as illustrated by the scatter in the velocity measurements.  We believe this discrepancy 
is due mainly to residual uncertainties in the FUSE pipeline wavelength calibration
and small imperfections in our alignment of the spectral segments.

Holberg et al. (1998) reported the detection of interstellar OI (1302.169\AA) and 
SiII (1190.416,1260.442\AA) in the IUE echelle spectrum of GD71 and determined a weighted 
mean velocity of +23.2$\pm$2.5kms$^{-1}$ for these features. Based on observations of objects with 
independently determined radial velocities, the absolute wavelength scale of IUE echelle data 
processed with NEWSIPS is found to be accurate to 3-5kms$^{-1}$ (Holberg et al. 1998). The nominal 
absolute calibration of the FUSE wavelength scale is considered to be limited to $\sim$15kms$^{-1}$ 
by uncertainty in the position of the target within the LWRS aperture (e.g Oegerle et al. 2005).
We have determined a weighted mean velocity of -0.3$\pm$0.6kms$^{-1}$ for the low ionization lines 
of CII (1036.337\AA) and OI (1039.230\AA) which lie within the wavelength range covered by the LiF 
1a segment (987.1-1082.3\AA) and almost certainly have the same origin as the lines reported by Holberg
et al. Since these results indicate an offset in the FUSE data of $\approx$-20kms$^{-1}$ with respect 
to the more robust wavelength scale of the IUE echelle spectrum in subsequent discussion we apply
 a systematic shift of +23.5$\pm$2.6kms$^{-1}$ to the velocities determined from the former.

Unfortunately, high resolution spectroscopic measurements of the core of the H-$\alpha$ line formed 
in atmosphere of GD71 indicate that the stellar photosphere has a similar velocity (e.g. the weighted mean of 
the measurements of Maxted et al. 2000 is +30.0$\pm$1.5kms$^{-1}$) to the interstellar medium (ISM) along 
this line of sight (+23.2$\pm$5.6kms$^{-1}$ allowing for a potential uncertainty of 5kms$^{-1}$ in the absolute 
calibration of the IUE echelle data). Thus radial velocities alone are not a suitable method for distinguishing 
between a photospheric and an interstellar origin for absorption features detected in this particular FUSE dataset. 
Nevertheless, when the comparatively high effective temperature of GD71, the low ionization
states of the progenitor species and previously published results based on FUSE observations of DA 
white dwarfs (e.g. Lehner et al. 2003) are taken into account the vast majority of lines listed in Table
2 can be attributed to the ISM. 

\begin{table}
\centering
\begin{minipage}{120mm}
\caption{Results of homogeneous model fitting proceedure.}
\begin{tabular}{@{}clcc@{}}
\hline
Star    &   T$_{\rm eff}$    & log g   &  [P/H] \\
\hline
GD71         & 32625$^{+15}_{-25}$ & 7.89$^{+0.01}_{-0.01}$ & -8.57$^{+0.09}_{-0.13}$  \\ 
RE J1918+595 & 32030$^{+95}_{-95}$ & 7.79$^{+0.03}_{-0.04}$ & -8.70$^{+0.23}_{-0.37}$  \\
RE J0605-482 & 34370$^{+190}_{-175}$ & 7.79$^{+0.06}_{-0.05}$ & -8.36$^{+0.14}_{-0.19}$  \\
\hline
\end{tabular}
\end{minipage}
\end{table}

\subsection[]{High ionization phosphorus lines in the FUSE spectrum of GD71}

\begin{figure*}
\vspace{115pt}
\includegraphics{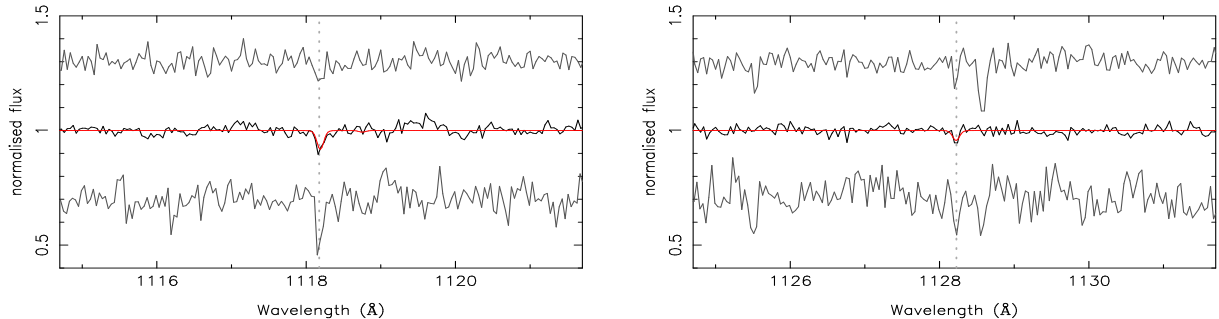}
\caption{Sections of the normalised FUSE spectra of REJ1917+599 (top), GD71 (middle) and REJ0605-482 
(bottom) showing the resonance lines of PV (1117.977,1128.008\AA) which we attribute to the stellar 
photospheres. Photospheric SiIV (1128.334\AA) and interstellar FeII (1125.448\AA) lines are also seen in 
the top and bottom datasets. The best fit model representation of the GD71 data is overplotted (red).}
\end{figure*}

A closer inspection of the coadded FUSE spectrum has also revealed several weak features arising from
species of relatively high ionization states e.g. the resonance lines of PV at 1117.977\AA\ and 
1128.008\AA. The velocities of these transitions are identical within the measurement uncertainties so 
they probably have a common origin (see Table 2). Although these lines are coincident in velocity 
space with a number of features attributable to the ISM, residual uncertainties in the wavelength calibration 
of this dataset and the similarity between the velocities of the ISM along this line of sight and the atmosphere
of GD71, as discussed above, mean that an interstellar, circumstellar or photospheric origin 
cannot be ruled out at this stage. Nevertheless, other considerations indicate the former interpretation 
is unlikely. For example, the ionization stages of the progenitor ions are higher than those typically found 
in the ISM. The PIII 1003.600\AA\ line arises from the 2P$^{o}$-2S multiplet and has a lower level 0.07eV 
above the ground state. Furthermore, in a solar mixture P is less abundant by a factor 1000 than C, N or O.

To examine the two other possible origins in more detail, we have acquired from MAST the FUSE data 
products for two further white dwarfs with similar effective temperatures and surface gravities to 
GD71. Vennes et al. (1997) derive T$_{\rm eff}$=33000K, log g=7.90 for REJ1918+595, while Marsh et al. (1997),
Finley et al. (1997) and Vennes et al. (1997) derive T$_{\rm eff}$=33040K, log g=7.80, T$_{\rm eff}$=35332K,
log g=7.84 and T$_{\rm eff}$=35600K, log g=7.76 respectively for RE J0605-482. FUV observations of RE 
J1918+595 and RE J0605-482, in TTAG mode and in the LWRS configuration were obtained on 2002/05/07 and 
2002/12/31 respectively. 

For each of these two stars we have constructed a co-aligned and co-added the FUSE spectrum in 
the manner described above. In each dataset we have determined the velocities of features attributable to 
the ISM, e.g. NI (1134.165, 1134.415, 1134.980\AA), FeII (1144.938\AA) and the stellar photosphere
e.g. SiIII (1108.368, 1109.957, 1113.219\AA) and SiIV (1122.485, 1128.340\AA), where the rest 
wavelengths of the Si lines are determined from the predictions of appropriate non-LTE model 
atmospheres. The weighted mean velocities of these interstellar and photospheric lines in the spectrum
of RE J1918+595 are +23.5$\pm$0.3kms$^{-1}$ and +59.8$\pm$0.9kms$^{-1}$ respectively. In the spectrum 
of RE J0605-482 these features have weighted mean velocities of +18.2$\pm$0.7kms$^{-1}$ and +53.1$\pm$
2.8kms$^{-1}$ respectively. Thus in the spectra of these two objects, the ISM and the 
photosphere have quite distinct velocities. 

Scrutiny of the data for RE J1918+595 and RE J0605-482 reveals the presence of PV resonance 
lines, at weighted mean velocities of +55.3$\pm$1.5kms$^{-1}$ and +53.7$\pm$0.6kms$^{-1}$ respectively.
A previous high resolution UV study has revealed circumstellar features along the lines of sight to nine 
out of a sample of 23 hot DA white dwarfs (Bannister et al. 2003). In the vast majority of 
these 9 cases the difference in the velocities of the circumstellar and interstellar features was found to
be less than 10kms$^{-1}$. The difference in velocity between the PV transitions and the ISM
lines in the FUSE spectra of RE J1918+595 and RE J0605-482 are 31.8$\pm$1.5kms$^{-1}$ and 35.5$\pm$3.0
kms$^{-1}$ respectively, arguing against a circumstellar origin. Indeed, given the scatter typically seen
in our measurements, the velocities of the PV lines strongly support a photospheric interpretation. In
Figure 1 we show sections of the FUSE spectra of GD71, RE J1918+595 and RE J0605-482 centered on the PV 
resonance transitions. For the purposes of this plot we have applied a shift to the wavelength scales to
co-align the 1117.977\AA\ line in the three datasets.

On the weight of the above evidence, we conclude that the most likely origin of the high ionization
P lines identified in the FUSE spectrum of GD71 is the stellar photosphere. Accepting this, we 
now use model atmosphere calculations to place constraints on the abundance of photospheric phosphorus 
in these three DAs.

\section[]{The model atmosphere calculations}

The models utilised in this work have been generated with the latest versions of the plane-parallel,
hydrostatic, non-LTE atmosphere and spectral synthesis codes
TLUSTY (v200; Hubeny 1988, Hubeny \& Lanz 1995, Hubeny \& Lanz private comm.) and SYNSPEC (v48; Hubeny et al.
1994). We have employed state-of-the-art model ions of H and P. The HI 
ion incorporates the 8 lowest energy levels and one superlevel extending from n=9 to n=80, where the
dissolution of the high lying levels was treated by means of the occupation probability formalism of Hummer \& 
Mihalas (1988), generalised to the non-LTE situation by Hubeny, Hummer \& Lanz (1994).
The PIV and PV model ions were developed by Lanz \& Hubeny (2003) for the study of O-stars and full 
details may be found therein. The PIII ion, which incorporates the 9 lowest levels, was constructed in a 
similar manner, except the online version of AUTOSTRUCTURE\footnote{http://random.ivic.ve/autos/} was used to estimate oscillator strengths 
where experimental values 
were unavailable. All calculations were carried out under the assumption of 
radiative equilibrium and incorporated a full treatment of line blanketing effects. 
During the calculation of the model structure the lines of the Lyman and Balmer series were treated by
means of an approximate Stark profile but in the spectral synthesis step, detailed profiles for the Lyman 
lines were calculated from the Stark broadening 
tables of Lemke (1997). 

%\subsection{Homogeneous modelling}

Numerous estimates of the effective temperature and surface gravity of GD71 are 
available in the literature e.g. Marsh et al. (1997), Finley et al. 
(1997) and Vennes et al. (1997) derive T$_{\rm eff}$=32008K, log g=7.70, T$_{\rm eff}$=32747K,
log g=7.68 and T$_{\rm eff}$=33000K, log g=7.88 respectively by fitting synthetic profiles to the 
observed Balmer lines. Hence we generated a grid of H+P models spanning the 
ranges T$_{\rm eff}$=30000-35000(2500)K, log g=7.5-8.5(0.5) and [P/H]=-12 -- -7.0(1.0). 
Our general spectral analysis techniques have been described at length in previous publications.
so only aspects specific to the current work are detailed here. For computational speed, only the 
regions of the co-added, optimal S/N, dataset which include the Lyman lines and the locations of
the phosphorus transitions predicted to be most prominent in this effective temperature and surface
gravity range were incorporated into the spectral fitting process (e.g. 930-990\AA, 1000-1050\AA\,
 1116-1120\AA\ and 1126-1130\AA). During the analysis, each of these sections was assigned an 
independent normalisation 
parameter to circumvent any residual errors in the shape of the continuum flux. All errors quoted 
here are $1\sigma$ unless stated otherwise but as these are derived formally from the fitting 
process, they may underestimate the true uncertainties. The parameters of our best fitting homogeneous
model representations of the FUSE spectra of all three white dwarfs are given in Table 3. We note
that the energy distribution of the GD71 model is indistinguishable from that of a pure-H 
composition from 10-10000\AA, except in a few narrow regions which include P lines.

\section{Discussion}

\subsection{Photospheric phosphorus in DA white dwarfs}

Bruhweiler (1984) reported the possible detection of NV, CIV and SiIV resonance lines in a single IUE 
spectrum of GD71 but concluded that these were most likely formed in a halo region around the star. A 
subsequent examination of a coadded dataset derived from the three existing IUE echelle spectra  
failed to confirm the existence of these features at the 30 m\AA\ equivalent width level and instead
detected only the strong resonance lines of interstellar OI and Si II (Holberg et al. 1998). Hence the 
current work represents the first probable detection of a heavy element in the atmosphere of the 
flux standard star GD71.

Phosphorus was first revealed in the atmospheres of the hot DA white dwarfs G191-B2B and MCT0455-2812 
by far ultraviolet ORFEUS observations (Vennes et al. 1996). Subsequently, photospheric phosphorus 
has been reported in the FUSE spectra of a significant number of other hot (T$_{\rm eff}$$\simgreat$40000K) 
hydrogen rich degenerates (e.g. GD246, Feige 55, Wolff et al. 2001 and  RE1032+532, Dupuis et al. 2004).
However, to date, phosphorus has been detected in the atmospheres of only two DAs with T$_{\rm eff}$
$<$40000K, GD394 (T$_{\rm eff}$$\approx$39500K; Chayer et al. 2000) and GD659 (T$_{\rm eff}$$\approx$36000K; 
Dupuis et al. 2004). The white dwarfs analysed in this paper, therefore, represent the three coolest 
DAs in which photospheric phosphorus has yet been found. 

\subsection{Possible mechanisms for photospheric phosphorus in these white dwarfs}

While the presence of a close cool companion can lead, through accretion of wind material, to heavy 
element contaminatation in a white dwarf atmosphere (e.g. V471 Tauri; Sion et al. 1997) it seems 
rather unlikely that this is the source of the phosphorus observed in these stars. For example, 
Maxted et al. (2000) find no evidence of radial velocity variations in GD71, while Dobbie et al. 
(2005) have used near-IR spectroscopy to set a limit of M$<$0.072M$_{\odot}$ on the mass of any such 
companion. Furthermore, 2MASS data provides no evidence of a near-IR excess to either REJ1917+488
or REJ0605-482 (e.g. Skrutskie et al. 1997). While the presence of a close substellar companion to any
of these stars cannot be excluded, recent work by Farihi et al. (2005) argues against this. 

Alternatively, equilibrium radiative levitation calculations indicate that observable quantities of 
phosphorus should be supported in the photospheres of typical DA white dwarfs to T$_{\rm eff}$$\approx$32500K 
(Vennes et al. 1996). Indeed, the phosphorus abundances we measure for these three stars seem to be 
satisfyingly consistent with the theoretical values (see their Figure 4). At face value, the lowest and 
largest abundance is measured in the coolest and hottest star respectively. However, the results of 
a preliminary study of the FUSE spectra of 28 hot DA white dwarfs indicate that this agreement might be
simply fortuitous. In many objects with 35000K$\simless$T$_{\rm eff}$$\simless$55000K, phosphorus is found 
to be underabundant, often strongly, with respect to the predictions (Dupuis et al. 2004). The abundance of
an element, in equilibrium, is generally depth dependent (e.g. Schuh 2005). The predictions of Vennes
et al. (1996) correspond to the phosphorus abundance at the Rosseland photosphere ($\tau_{\rm Ross}$$\approx$2/3),
which may differ significantly from that in the line formation region from which the observed abundances are
determined. Therefore, to appraise the r\^{o}le of radiative levitation in more detail we have constructed a 
small number of fully self-consistent H+P model atmospheres. In these calculations we have used information 
on the ionization fraction of phosphorus and the detailed radiation field gleaned 
from SYNSPEC to determine the effective gravitational downward forces and the upward radiative forces on 
phosphorus ions as a function of depth in the atmosphere. An estimate of the phosphorus 
abundance at each depth point has been obtained by equating these forces (e.g. Chayer et al. 1995b). The 
estimates have been fed back into TLUSTY and an updated model structure determined. The entire process has
been repeated until the abundances converge to $\simless$3\% throughout most of the atmosphere. In the 
uppermost layers this restiction has been relaxed to $\simless$10\% but has no significant impact on the 
emergent spectrum.

The results of these calculations consolidate previous predictions that detectable levels of phosphorus 
remain supported in the atmospheres of typical DA white dwarfs at the effective temperatures examined here. 
However, the metal absorption lines predicted by our non-LTE models are considerably stronger than those 
observed in the FUSE data. For example, if we attempt to fit our homogeneous grid to the emergent spectra 
from the self-consistent calculations with effective temperatures and surface gravities appropriate to RE 
J1918+595 and GD71, we derive [P/H]$\approx$-7.3. If instead we fit the self-consistent models to the 
observed phosphorus features only, fixing the effective temperature at the values determined in \S 3.1, 
we find the element abundance profile appropriate to log g=8.43, 8.44 and 8.45 provides the best representation 
of these data, for RE J0605-482, GD71 and RE J1918+595 respectively. These gravities are greater by $\sim$0.6 dex than 
those estimated from the Lyman line series. The apparent discrepancy between the results from our 
modelling and the predictions of Vennes et al. (1996) likely stems from the abundance of phosphorus 
at the Rosseland photosphere, log$_{10}$ (model column mass depth) $\sim$\thinspace-2, decreasing rapidly with effective 
temperature at T$_{\rm eff}$$\simless$35000K, while in the P line forming region, log$_{10}$ (model column mass depth)
$\sim$\thinspace-3.5, maintaining a greater value.

Interestingly, in their analysis of the EUVE spectra of 26 DA white dwarfs (T$_{\rm eff}$
$\simgreat$40000K), where a grid of self-consistent stratified models was compared by eye to the data, 
allowing effective temperature and surface gravity to, in effect, vary freely, Schuh et al. (2002) found, 
more often than not, the calculation which best represented the observed energy distribution of a white dwarf 
had a larger surface gravity ($\sim$0.4 dex) than the value determined for the star via Balmer line fitting. 
Schuh (2005) has estimated that the complexities in accurately calculating the radiative force on
an element in a stellar photosphere, e.g. due to the opacities of other metals, can result in 
uncertainties of up to 2 dex in the abundances derived at some depth points in their calculations. We
note that the three white dwarfs analysed here are more metal poor and somewhat cooler than those typical 
of her sample, lying in a range where non-LTE models replicate the structure of the stellar photosphere 
more satisfactorily (Barstow et al. 2003b). Indeed, test calculations reveal that the inclusion of 
additional metals, e.g. C and Si, has no significant impact on the P abundance profile. 
However, robust and comprehensive atomic data for P are not available from TOPBASE (Cunto \& Mendoza 1992)
so shortcomings in the current model ions and linelists for this element may counter any advantage here. 

Of course the balance between radiative levitation and gravitational settling can be disturbed by 
massloss. From a series of exploratory time dependent calculations involving Si, Chayer et al. (1997) 
find that at moderate mass loss rates (10$^{-16}$$\simless$$\dot{\rm M}$$\simless$10$^{-14}$M$_{\odot}$ yr$^{-1}$),
the reservoir of this element is exhausted after a few$\times$10$^{4}$ years, after which it is depleted
rapidly from the stellar photosphere (few$\times$10$^{5}$ yrs) to levels well below the detection 
threshold of the FUSE data. At very low rates of massloss (e.g. $\dot{\rm M}$$\simless$10$^
{-18}$M$_{\odot}$ yr$^{-1}$) the Si abundance 
profile is essentially that predicted by equilibrium radiative levitation theory up to 
the end of their calculations. The mere detection of phosphorus in these three stars would appear to argue 
against massloss at the larger rates discussed above. Nevertheless, it seems plausible that this process may 
be operating at low levels in these stars and that it is the slow exhaustion of an underlying reservoir of 
this element which results in observed abundances below the equilibrium radiative levitation prediction. 

\section*{Acknowledgments}
PDD and MBU are supported by PPARC. JBH wishes to acknowledge support from NASA ADP grant NNG05GC46G.
Based on observations made with the NASA-CNES-CSA Far Ultraviolet 
Spectrscopic Explorer. FUSE is operated for NASA by the John Hopkins University under NASA contract 
No. NAS5-32985. We express our thanks to the anonymous referee for a very prompt and useful report.

\bsp

\label{lastpage}

\end{document}